\documentclass[12pt]{report}
\usepackage[utf8]{inputenc}
\usepackage{blindtext}
\usepackage[letterpaper, total={8.5in, 11in}, margin=2in]{geometry}
\usepackage{ragged2e}
\usepackage{blindtext}
\usepackage{graphicx}
\usepackage{amsmath}
\usepackage{amssymb}
\usepackage{gensymb}

\date{}

\begin{document}
%\vspace*{\fill}
\centering{\huge The Deployment of IRS in UAV-Empowered 6G Networks\\
\vspace{30pt}
\large Mobasshir Mahbub, Raed M. Shubair}
%\vspace*{\fill}
\newpage

\RaggedRight{\textbf{\Large 1.\hspace{10pt} Introduction}}\\
\vspace{18pt}
\justifying{\noindent The last two decades have witnessed an exponential growth and tremendous developments in wireless technologies and techniques, and their associated applications, such as those reported in [1]–[30].\par
The data rate will immensely increase attributed to the prevalence of increased users in hotspot circumstances. To solve the challenge, 6G (sixth generation) systems must migrate to several sophisticated network enhancement techniques e.g. IRS [31], [32], THz communication [33], etc.\par
UAV technology is regarded as a critical component of 6G mobile networks [34]. In comparison to typical terrestrial cellular connectivity, UAVs can swiftly facilitate connections in hotspot settings due to their great mobility. It is also cost-effective in regions with insufficient coverage, such as remote or hilly locations. As a result, the design of 6G networks may transition from the stationary terrestrial infrastructure model to aerial mobile connectivity.\par
IRS [35], [36], also referred to as a metasurface [37], is a newly developed engineered material whose radiation characteristics, such as reflection, phase, absorption, and refraction can be electronically modified in real-time. Because these surfaces are low-cost to produce, they may be distributed globally, giving an unparalleled opportunity to manipulate the wireless multipath scenario both indoors and outdoors. As a result, IRS offers up an entirely new avenue in mobile communications research, shifting the emphasis from countering multipath to creating it. Indeed, numerous studies have recently proven that IRS-assisted solutions may increase the coverage, capacity, energy, spectral efficiency of contemporary mobile networks dramatically. Fig. 1 visualizes the IRS-empowered UAV-assisted network.

\vspace{6pt}

\begin{figure}[h]
    \centering
    \includegraphics[height=9.0cm, width=11.5cm]{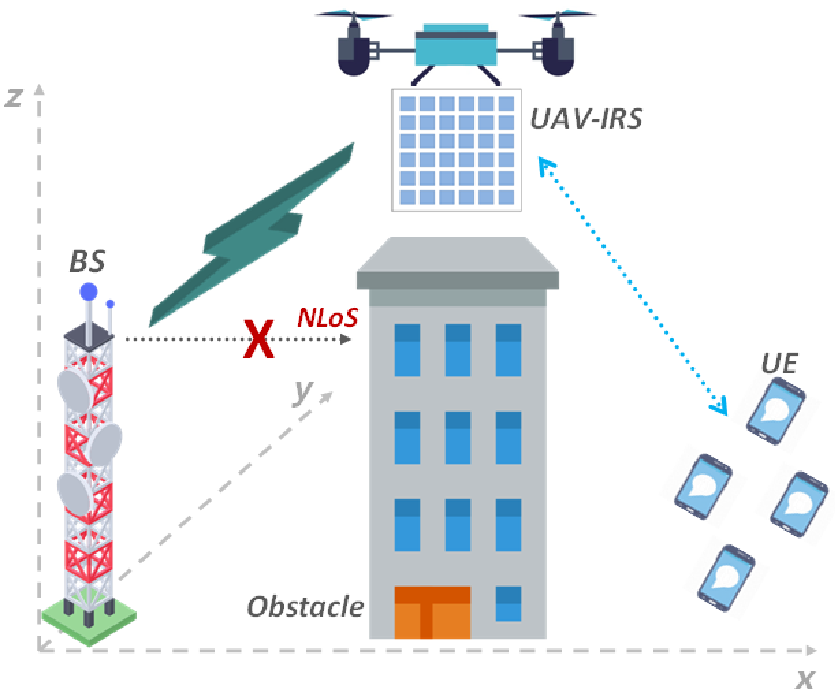}\\
    Fig. 1. IRS-empowered UAV-assisted wireless network
\end{figure}
UAVs are becoming increasingly used for relaying, data collection, secure communication, and information distribution. Because of their great agility and versatility for on-demand implementation, UAVs have a high likelihood of having line of sight (LoS) connections [34]. As a result, UAVs can be used as aerial communication mediums to augment the performance of current ground wireless transmission networks, such as mobile networks.\par
However, UAV communication confronts several problems, particularly in urban environments. One notable concern is obstruction created by ordinary things including buildings/ infrastructure, forests/trees, and human bodies that can exacerbate coverage and connection issues [38]. The movement of both the UAV and the UE (user equipment), on the contrary, causes excessive temporal and spatial changes in the non-stationary communication channels.\par
To solve these issues, the intelligent reflecting surface (IRS) technology [39] has recently been proposed to avoid impediments and improve connectivity in UAV systems, referred to as IRS-empowered UAV network systems [40]. IRSs are installed in the open network environment to aid the connection between UAVs and users, according to the IRS-assisted UAV architecture. A blocked i.e. non-line of sight (NLoS) transmission channel can be mitigated by deploying the IRS enabling several LoS links, which considerably minimizes channel attenuation.\par
Therefore, the work targeted minimizing the path loss and maximizing the achievable data rate in IRS-UAV-assisted networks.}
\vspace{18pt}

\RaggedRight{\textbf{\Large 2.\hspace{10pt} Relative Literature}}\\
\vspace{18pt}
\justifying{\noindent The paper in this section briefed several prior works and literature relative to IRS-empowered UAV communication. Since IRS-assisted wireless communication is an emerging topic and research is ongoing, during the literature review the authors find that there is a limitation of literature relative to path loss measurement in the IRS-UAV communication scenario.\par
Ma et al. [41] analyzed the potential deployment of IRS in cellular communications with UAVs that suffer from degraded signal strength. The work considered the implementation of IRS on walls that can be configured remotely by base stations to coherently transmit the reflected signal towards corresponding UAVs to enhance signal strengths at the user end. Al-Jarrah et al. [42] presented the outage probability and symbol error rate (SER) analysis of multi-layer UAV-empowered wireless communications supported by the IRS. The research of Pang et al. [43] overviewed the amalgamation of UAV and IRS, by illustrating the applications of IRS and the advantages of UAV and describing the advantages of incorporating them in a combined manner in the wireless network. Then, the work investigated case studies namely the UAV trajectory, the passive beamforming in IRS, and the transmit beamforming at the base station are jointly optimized. Jiang et al. [44] proposed and analyzed a three-dimensional (3D) stochastic geometry-based channel model incorporating multiple-input multiple-output (MIMO) for IRS-aided UAV communications. Wei et al. [45] considering the application of IRS in UAV-assisted orthogonal frequency division multiple access (OFDMA) transmission systems first derived the expression to analyze composite channels and proposed an approximation approach to set a lower and an upper bound for the considered problem. Mahmoud et al. [46] investigated the deployment of IRS in UAV-empowered communications networks aiming to enhance the coverage and improve the reliability in terms of spectral efficiency considering the Internet of Things (IoT) paradigm. Specifically, the work first derived tractable analytic expressions and then analyzed the ergodic capacity, achievable SER, and outage probability. Cao et al. [47] proposed and analyzed a 3D non-stationary MIMO channel model for an IRS-assisted UAV communications network. Shafique et al. [48] presented and analyzed a theoretical framework of an IRS-integrated UAV relaying system.}\par
\vspace{18pt}

\RaggedRight{\textbf{\Large 3.\hspace{10pt} Measurement Model}}\\
\vspace{12pt}
\RaggedRight{\textit{\large A.\hspace{10pt} Conventional UAV-Assisted Communication Model}}\\
\vspace{12pt}
\justifying{\noindent Contemplate a 3D geographic coverage region in which terrestrial base stations, UAVs, and user equipment are spanned in a 3D plane. A set of base stations B are deployed in the mentioned communication (or coverage) region along with U set of UAVs to enhance the coverage (in case of NLoS and obstacle-filled area) for a set of user N. Consider $u_i=(x_i^{uav},y_i^{uav},z_i^{uav})$ as the location (3D coordinates) of UAV $i \in U$ and $(x_n^{UE},y_n^{UE},z_n^{UE})$ as the coordinates denoting the location of user $n \in N$ Therefore, according to the ITU (International Telecommunication Union), the path loss is formulated as (Eq. 1) [49],}
\vspace{6pt}
\begin{equation}
PL= (\frac{4\pi f_c d_0}{c})^2 (\frac{d_i}{d_0 })^2 u_{NLoS}
\end{equation}
where $u_{NLoS}$ denotes the attenuation factor. c denotes the speed of light in $ms^-1$. $f_c$ indicates the carrier frequency in Hz. $d_0$ denotes the reference distance (free space). Here, $d_0=1$ m. $d_i= \sqrt{(x_n^{UE}-x_i^{uav})^2+(y_n^{UE}-y_i^{uav})^2+(z_n^{UE}-z_i^{uav})^2}$ indicates the distance between the UAV i and arbitrary user equipment (UE) located at $(x_n^{UE},y_n^{UE},z_n^{UE})$ coordinates. The received power thereby can be obtained as follows (Eq. 2),
\vspace{6pt}
\begin{equation}
p_r= \frac{p_t}{K_0 d_i^2 u_{NLoS}}
\end{equation}
\vspace{6pt}
where $p_t$ is the transmit power and $K_0=(\frac{4\pi f_c d_0}{c})^2$. The SNR (Signal to Noise Ratio) can be formulated by the following equation (Eq. 3),
\vspace{6pt}
\begin{equation}
S= \frac{p_r}{\sigma^2}
\end{equation}
\vspace{6pt}
where $\sigma^2 = -90$ dBm denotes the Gaussian noise power. The achievable data rate $(bits/s/Hz/m^2)$ by the user equipment can be determined by (Eq. 4),
\vspace{6pt}
\begin{equation}
R= log_2(1+\frac{p_r}{\sigma^2})
\end{equation}\par
\vspace{12pt}

\RaggedRight{\textit{\large B.\hspace{10pt} IRS-Embedded UAV-Assisted Communication Model}}\\
\vspace{12pt}
\justifying {\noindent Incorporating the IRS in UAV the far-field beamforming model for determining the path loss in the case of IRS-enhanced UAV-assisted communication model is (Eq. 5) [50], [51],
\vspace{6pt}
\begin{equation}
PL_{IRS} = \frac{64\pi^3(d_1 d_2)^2}{G_t G_r G M^2 N^2 d_x d_y \lambda^2 cos(\theta_t) cos(\theta_r) A^2}
\end{equation}
\vspace{6pt}
where $d_1=  \sqrt{(x^{BS}-x_i^{IRS} )^2+(y^{BS}-y_i^{IRS} )^2+(z^{BS}-z_i^{IRS} )^2}$ denotes the separation distance between the transmitter i.e. base station located at $(x^{BS},y^{BS},z^{BS})$ coordinates and IRS-embedded UAV located at $(x_i^{IRS},y_i^{IRS},z_i^{IRS})$ coordinates. $d_2=  \sqrt{((x_i^{IRS}-x_n^{UE} )^2+(y_i^{IRS}-y_n^{UE} )^2+(z_i^{IRS}-z_n^{UE} )^2}$
denotes the separation distance between the IRS and the UE. $G_t$ and $G_r$ are the gains of the transmitter and receiver and $G=\frac{4\pi d_x d_y}{\lambda^2}$  is the scattering gain. M and N are the numbers of transmission and reception elements of the IRS. $d_x$ and $d_y$ are the length and width of each scattering element. $\lambda$ denotes the wavelength of the transmitted signal. $\theta_t$ and $\theta_r$ are the transmitter and receiver angles with the IRS. $A$ denotes the amplitude relative to the reflection coefficient of unit cells of the IRS. Thereby, the received power at the UE can be formulated by (Eq. 6),
\vspace{6pt}
\begin{equation}
p_r^{IRS} = \frac{p_t G_t G_r G M^2 N^2 d_x d_y \lambda^2 cos(\theta_t) cos(\theta_r) A^2}{64\pi^3(d_1 d_2)^2}
\end{equation}\par
\vspace{6pt}
The SNR can be measured by the following formula (Eq. 7),
\begin{equation}
S^{IRS}= \frac{p_r^{IRS}}{\sigma^2}
\end{equation}\par
\vspace{6pt}
The achievable can be calculated by the following equation (Eq. 8),
\begin{equation}
R^{IRS}= log_2(1+\frac{p_r^{IRS}}{\sigma^2})
\end{equation}}\par
\newpage

\RaggedRight{\textbf{\Large 4.\hspace{10pt} Results and Discussions}}\\
\vspace{12pt}
\justifying{\noindent This section includes the measurement results based on the measurement model utilizing MATLAB-based simulation and incorporates discussions on the derived simulation result. The research considered low-altitude UAV since it utilized mmWave carrier.\par
Fig. 2 (a) and (b) show the path loss in terms of transmitter-receiver separation distance with 2D and 3D figures respectively in the case of conventional UAV-assisted communication networks. $f_c$ = 100 GHz, $(x_i^{uav},y_i^{uav},z_i^{uav})$ = (0, 0, 40) m, $(x_n^{UE},y_n^{UE},z_n^{UE})$ = (0-100, 0-100, 1.5) m, $u_{NLoS}$ = 23 dB indicate the simulation parameters.}\par
\vspace{6pt}

\centering
\includegraphics[height=8.0cm, width=10.0cm]{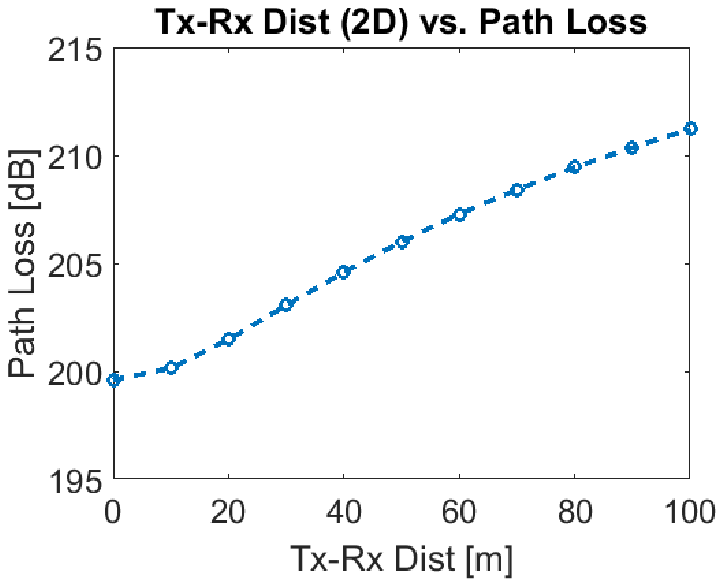}\par
\vspace{6pt}
(a)\par
\vspace{6pt}
\includegraphics[height=8.0cm, width=10.0cm]{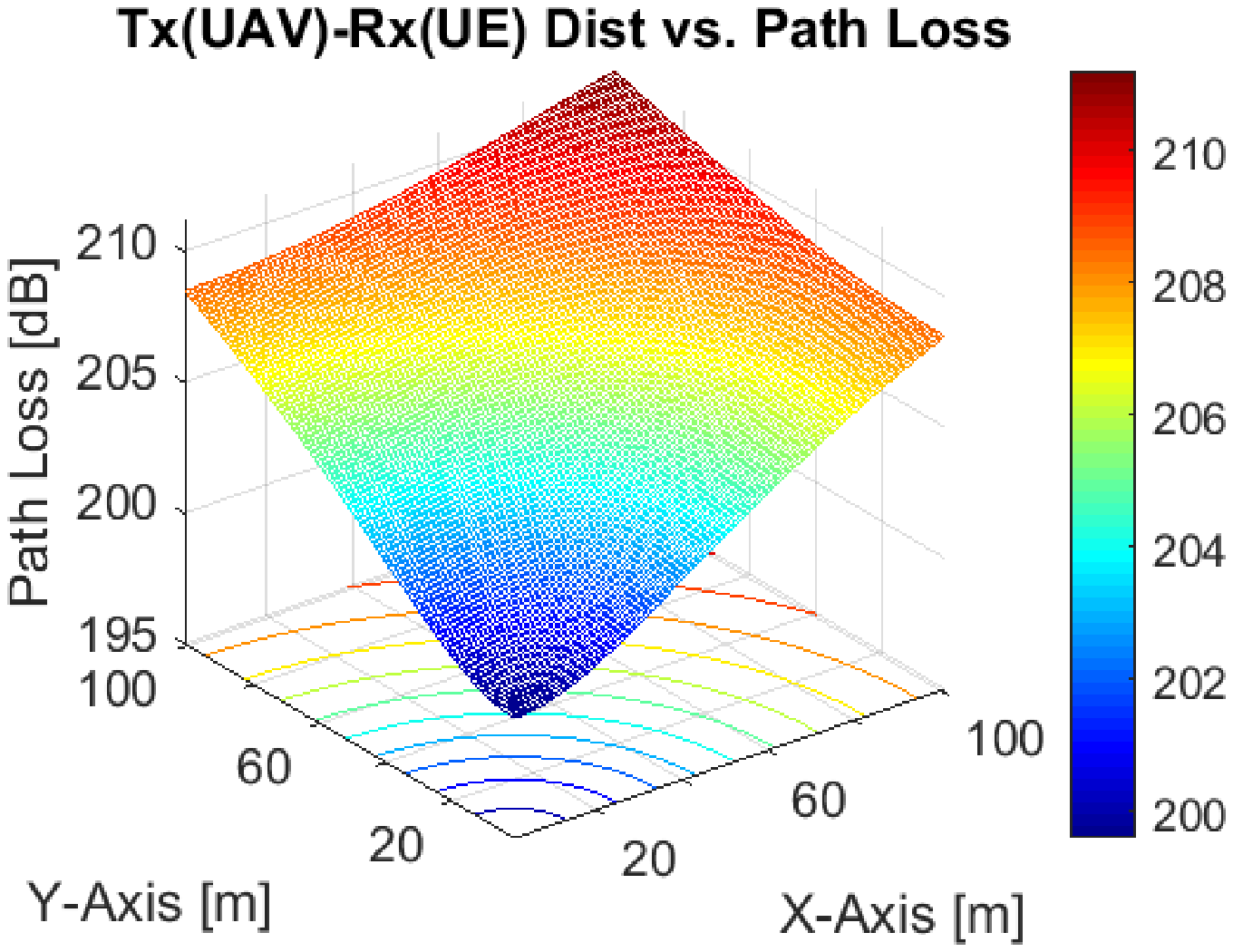}\par
(b)\par
\vspace{12pt}
Fig. 2. (a) Tx-Rx separation distance vs. path loss (2D); (b) Tx-Rx separation distance vs. path loss (3D)\par
\vspace{6pt}

\justifying{Fig. 3 (a) and (b) represent the achievable data rate in terms of transmitter-receiver separation distance with 2D and 3D figures respectively considering the conventional UAV-assisted communication model. For this measurement $p_t$ = 6W (other parameters are as same as Fig. 2).}
\vspace{6pt}

\centering
\includegraphics[height=8.0cm, width=10.0cm]{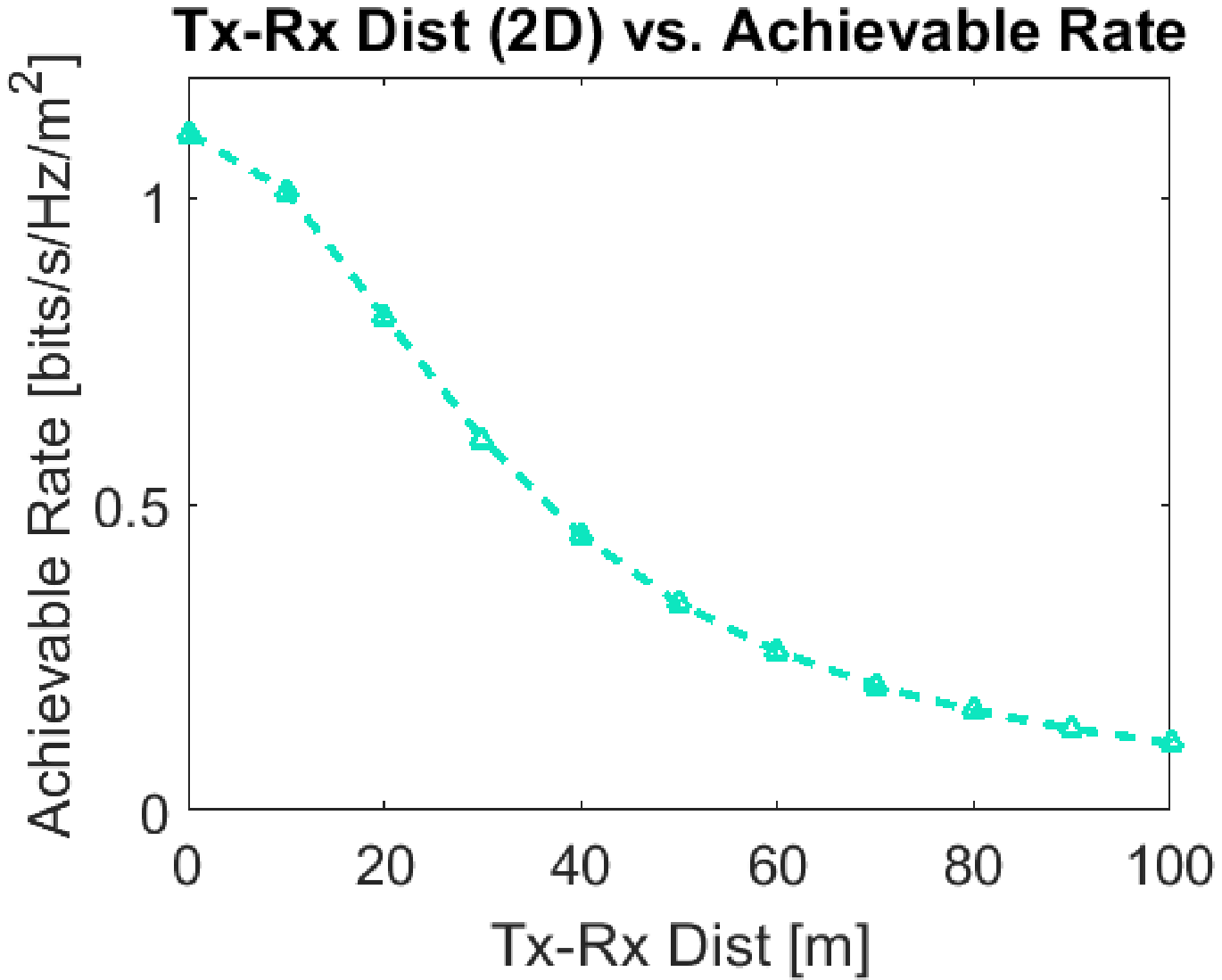}\par
(a)\par
\vspace{6pt}

\centering
\includegraphics[height=8.0cm, width=10.0cm]{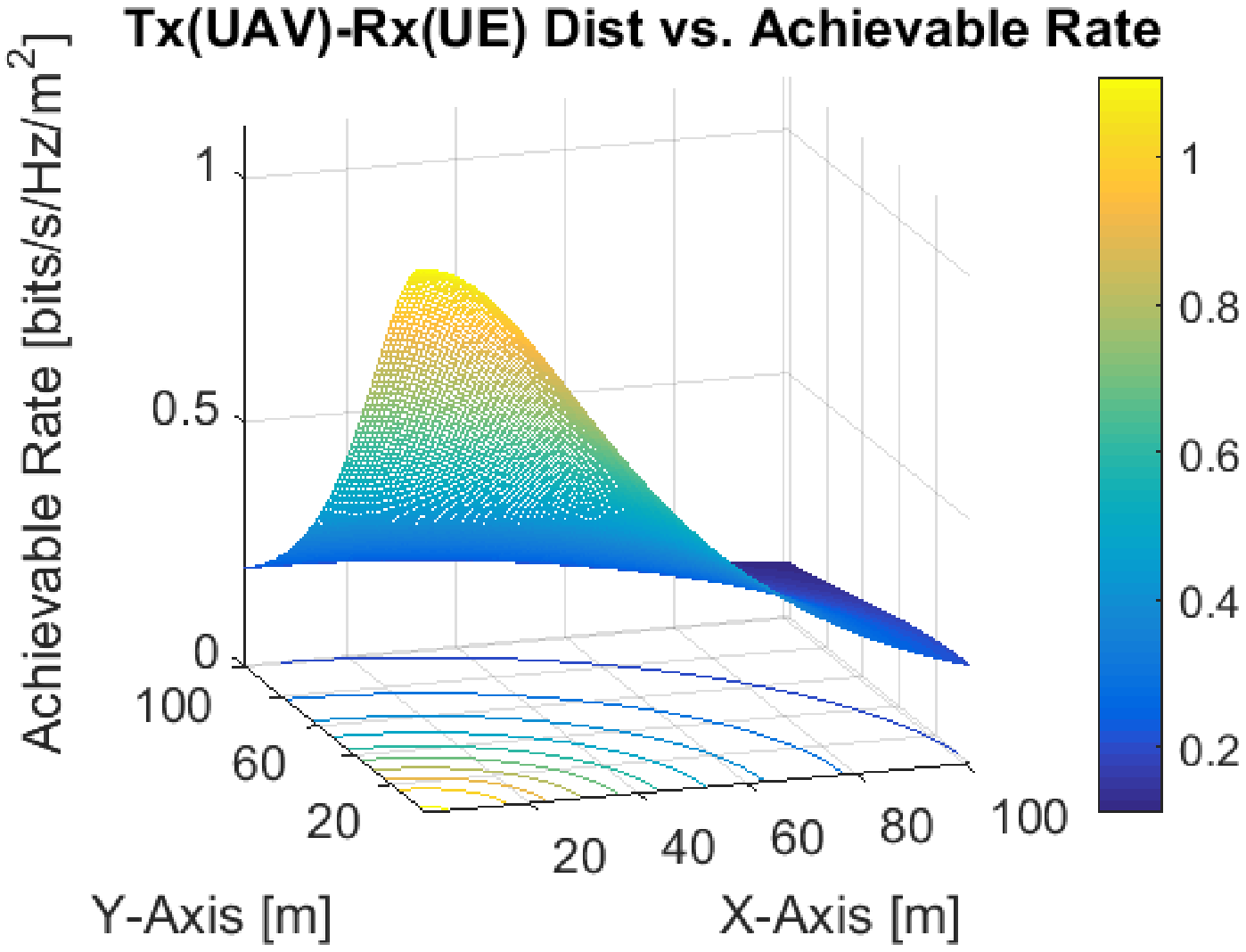}\par
(b)\par
\vspace{6pt}
Fig. 3. (a) Tx-Rx separation distance vs. achievable rate (2D); (b) Tx-Rx separation distance vs. achievable rate (3D)
\vspace{6pt}

\justifying{Fig. 4 (a) illustrates the path loss in terms of transmit-receive angle with a 2D figure for better realization and Fig. 4 (b) shows the path loss in terms of transmit-receive angle with 3D figures for $0\degree$ to $90\degree$ angles (from BS to IRS and IRS to UE) respectively in the context of IRS-empowered UAV communications model. $f_c$ = 100 GHz, $(x^{BS},y^{BS},z^{BS})$ = (0, 0, 8) m, $(x_i^{IRS} ,y_i^{IRS} ,z_i^{IRS})$ = (50, 50, 40) m, $(x_n^{UE},y_n^{UE},z_n^{UE})$ = (100, 100, 1.5) m, $G_t$ = 20 dB, $G_r$ = 20 dB, M = 100, N = 100, $d_x$ and $d_y = \lambda⁄2$, A = 0.9 (tx-rx angles are varied).}
\vspace{6pt}

\centering
\includegraphics[height=8.0cm, width=10.0cm]{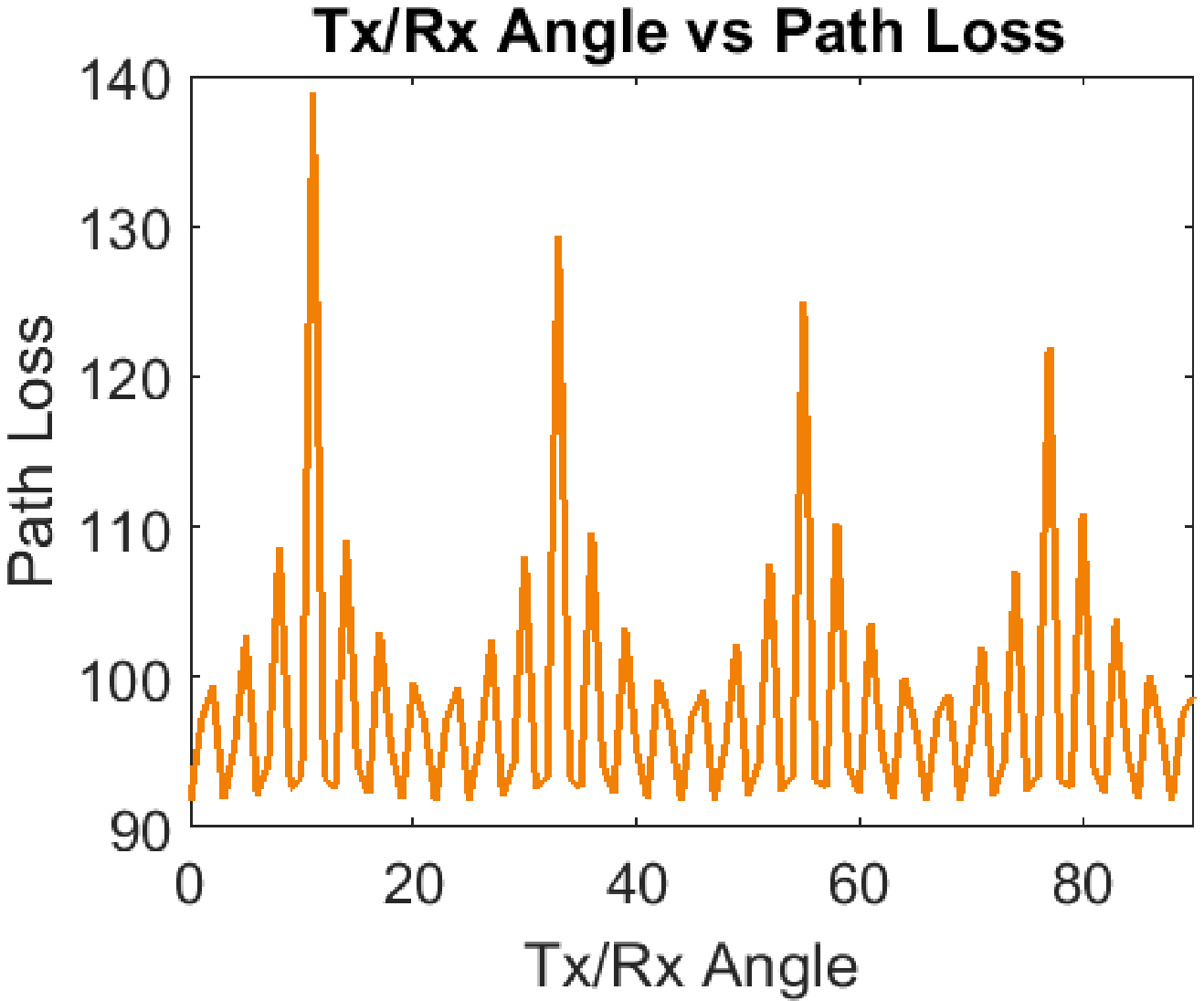}\par
(a)\par
\vspace{6pt}

\centering
\includegraphics[height=8.0cm, width=10.0cm]{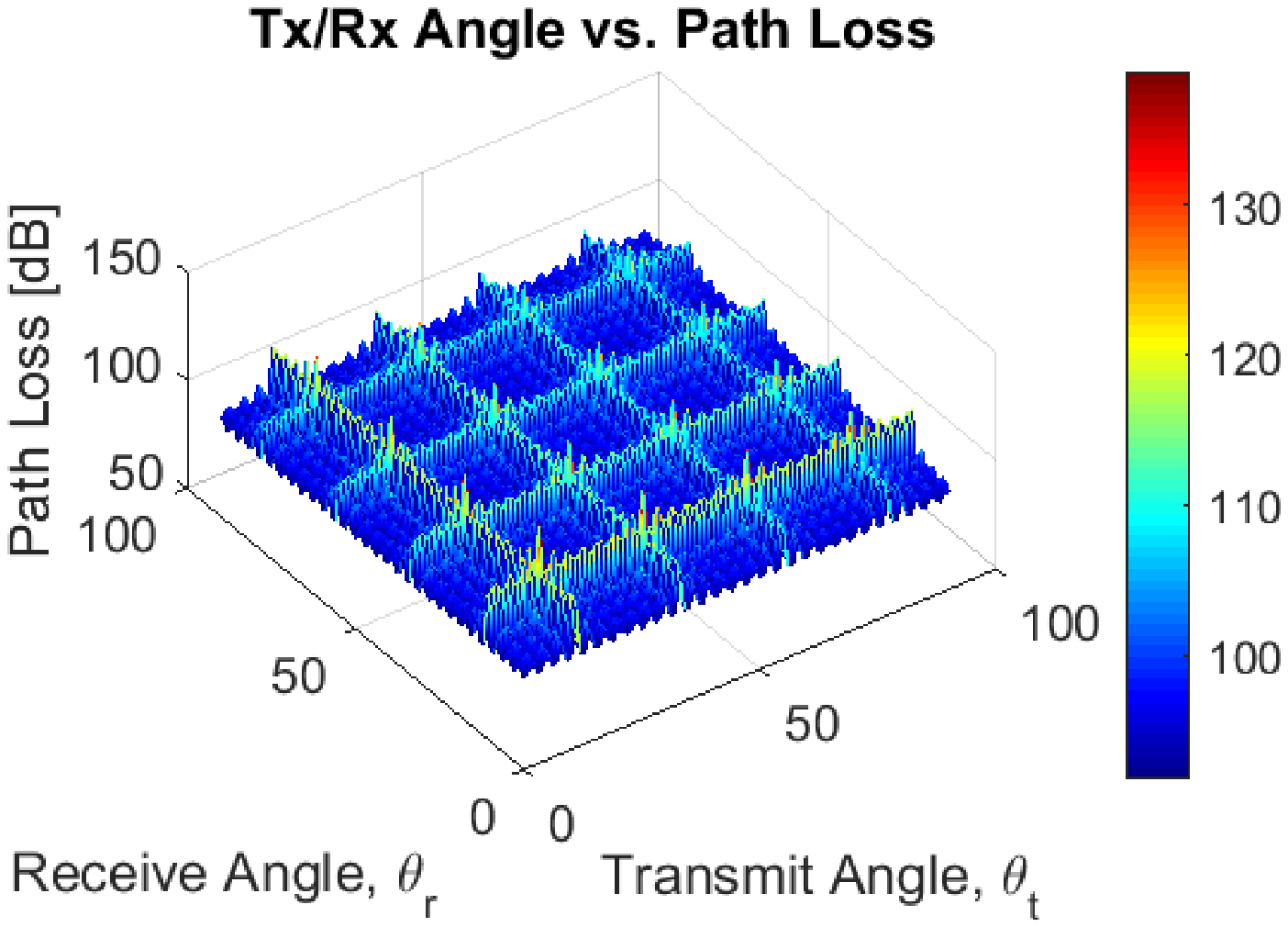}\par
(b)\par
\vspace{6pt}
Fig. 4. (a) Tx-Rx angle vs. path loss (2D); (b) Tx-Rx angle vs. path loss (for angle $0\degree$ to $90\degree$)
\vspace{6pt}

\justifying{Fig. 5 (a) and (b) visualize the path loss with top view and 3D figures in terms of transmitter-receiver separation distance (from BS to IRS and IRS to UE) in the case of IRS-empowered UAV communication networks. $\theta_t$  and $\theta_r = 45\degree$, $x_i^{IRS}$ and $y_i^{IRS}$ are varied only, (other parameters are as same as Fig. 4).}
\vspace{6pt}

\centering
\includegraphics[height=8.0cm, width=10.0cm]{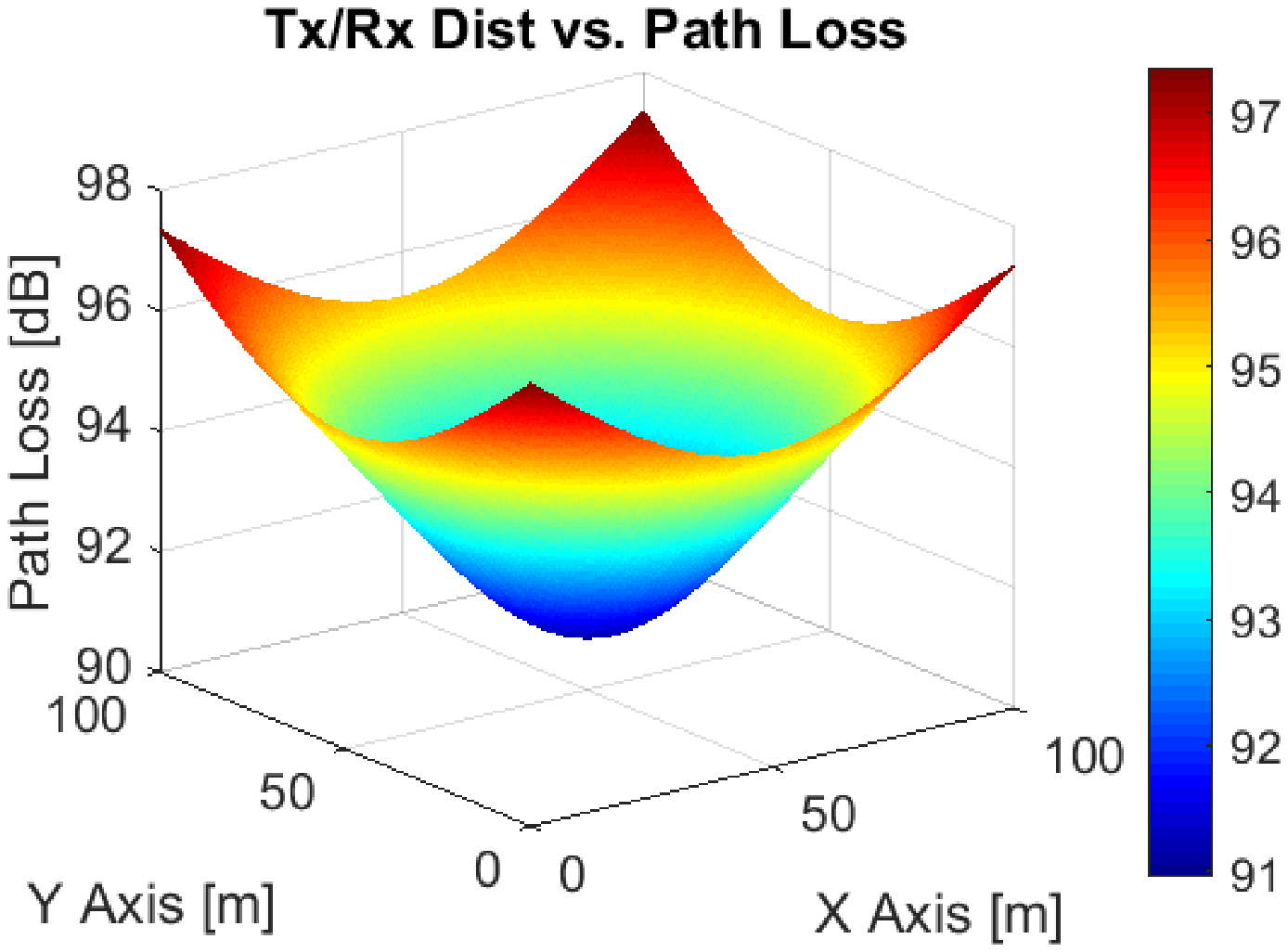}\par
(a)\par
\vspace{6pt}

\centering
\includegraphics[height=8.0cm, width=10.0cm]{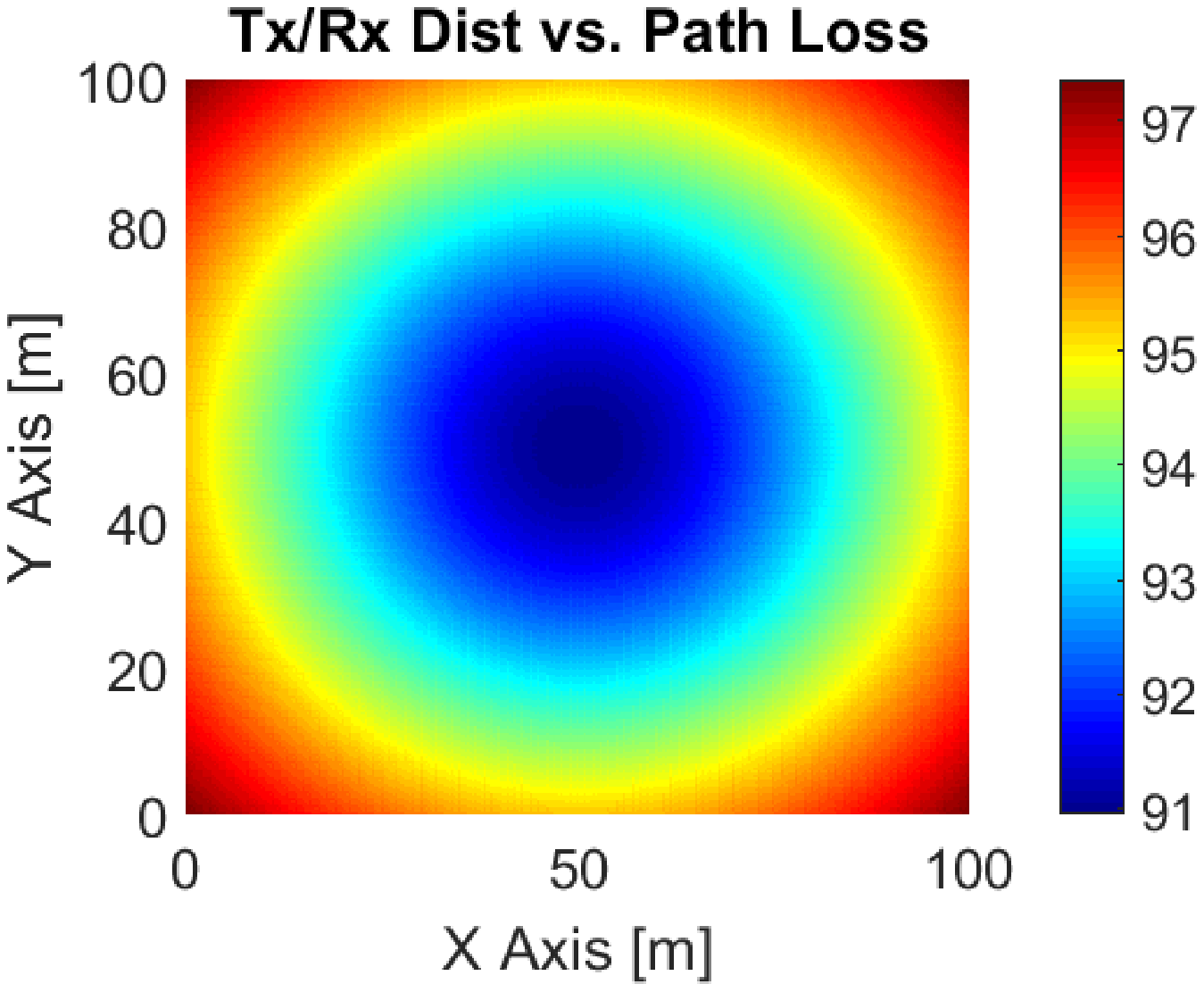}\par
(b)\par
\vspace{6pt}
Fig. 5. (a) Tx-Rx separation vs. path loss (top view); (b) Tx-Rx separation vs. path loss (3D)
\vspace{6pt}

\justifying{Fig. 6 represents the path losses in terms of the number of transmit-receive elements of IRS. M and N are varied only, (other parameters are as same as Fig. 4 and angles of Fig. 5).}
\vspace{6pt}

\centering
\includegraphics[height=8.0cm, width=10.0cm]{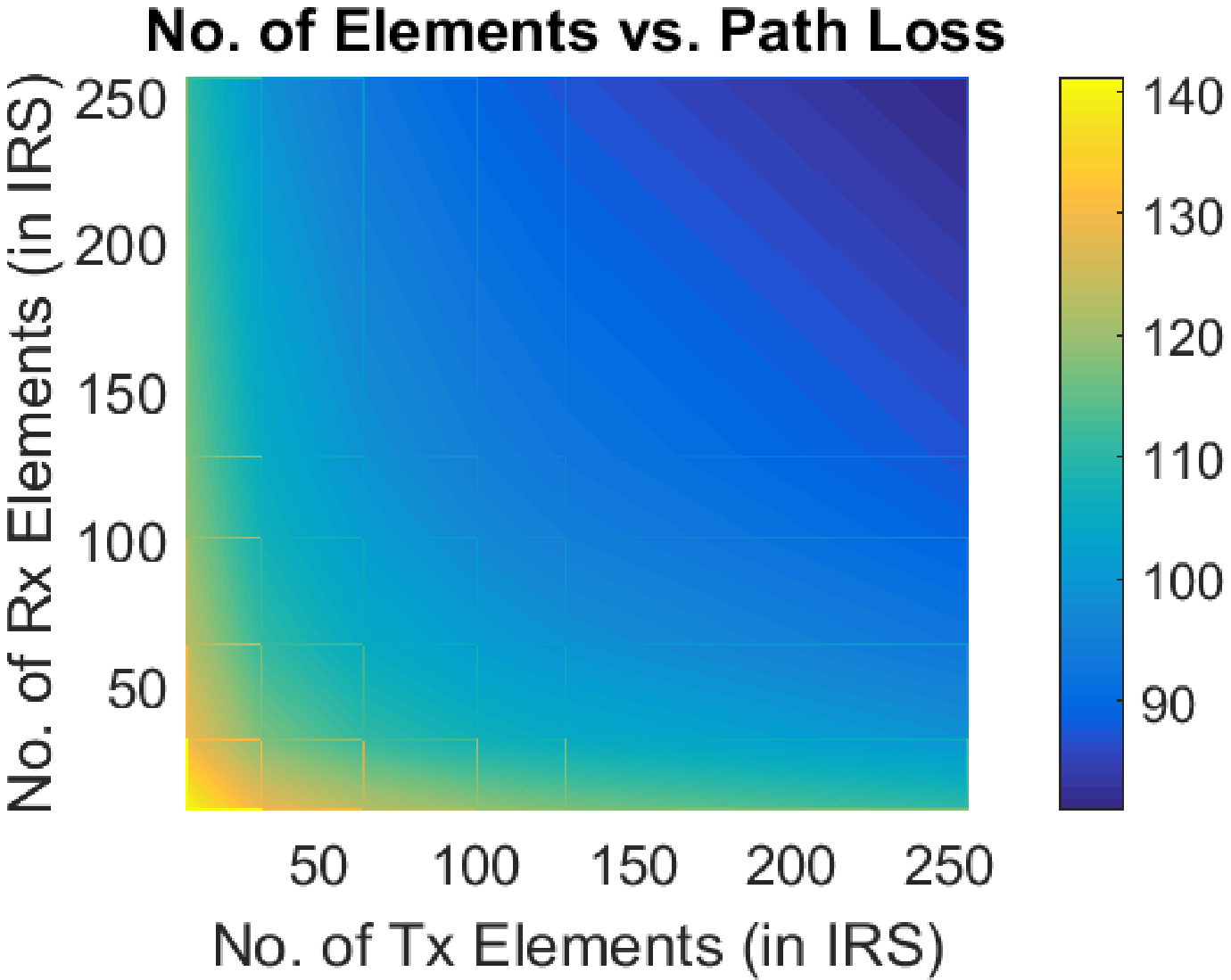}\par
\vspace{6pt}
Fig. 6. No. of Tx-Rx elements vs. path loss
\vspace{6pt}

\justifying{Fig. 7 (a) shows the achievable data rate in terms of transmit-receive angle with a top view figure and Fig. 7 (b) shows the path loss in terms of $0\degree$ to $90\degree$ transmit-receive angles in the context of the IRS-empowered UAV communications model. $p_t$ = 2W (parameters are as same as Fig. 4).}
\vspace{6pt}

\centering
\includegraphics[height=8.0cm, width=10.0cm]{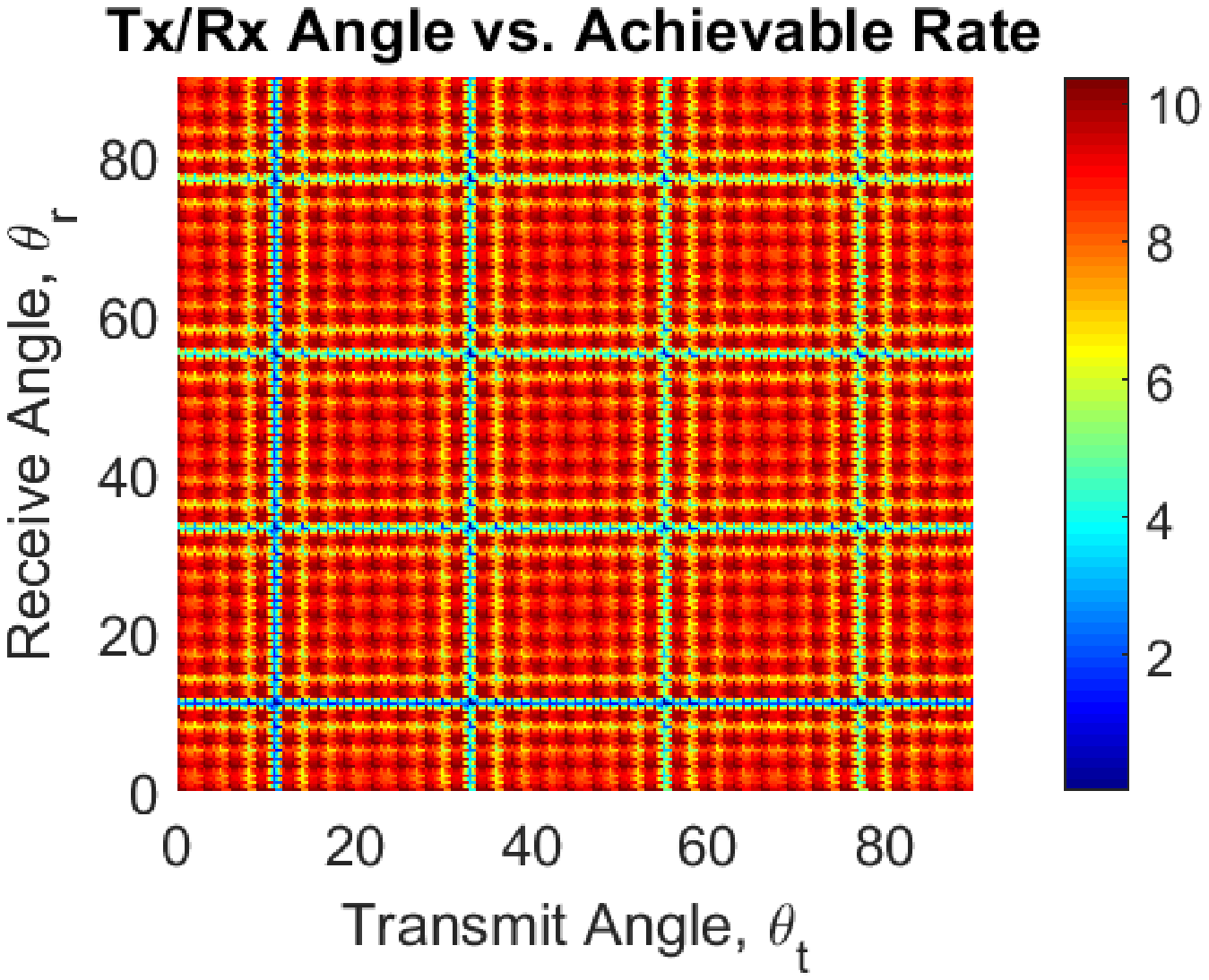}\par
(a)\par
\vspace{6pt}

\centering
\includegraphics[height=8.0cm, width=10.0cm]{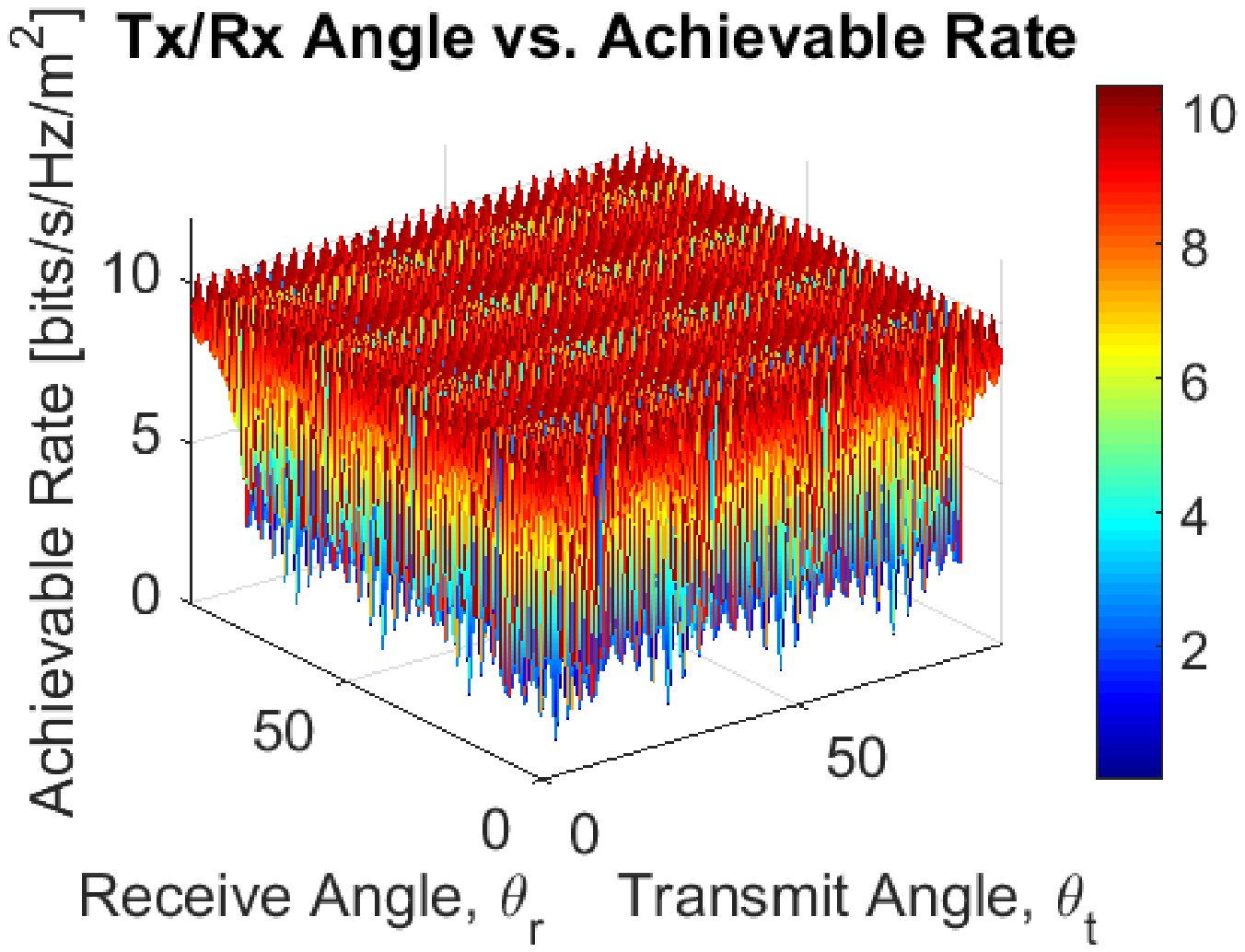}\par
(b)\par
\vspace{6pt}
Fig. 7. (a) Tx-Rx angle vs. achievable data rate (top view); (b) Tx-Rx angle vs. achievable data rate (for angle $0\degree$ to $90\degree$)
\vspace{6pt}

\justifying{Fig. 8 (a) and (b) represent the achievable data rate with top view and 3D figures in terms of transmitter-receiver separation distance in the case of IRS-empowered UAV communication networks. $p_t$ = 2W (parameters are as same as Fig. 5).}
\vspace{6pt}

\centering
\includegraphics[height=8.0cm, width=10.0cm]{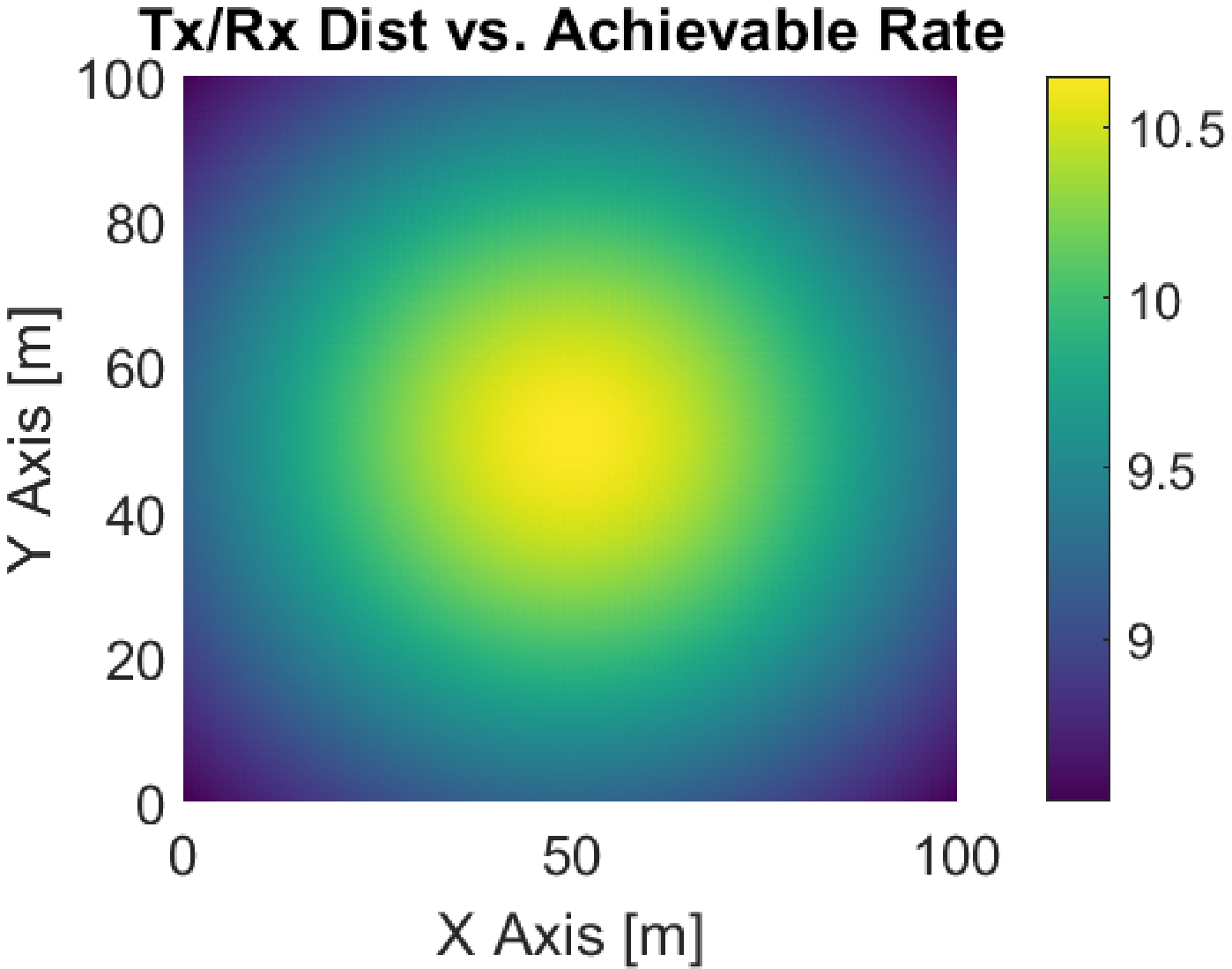}\par
(a)\par
\vspace{6pt}

\centering
\includegraphics[height=8.0cm, width=10.0cm]{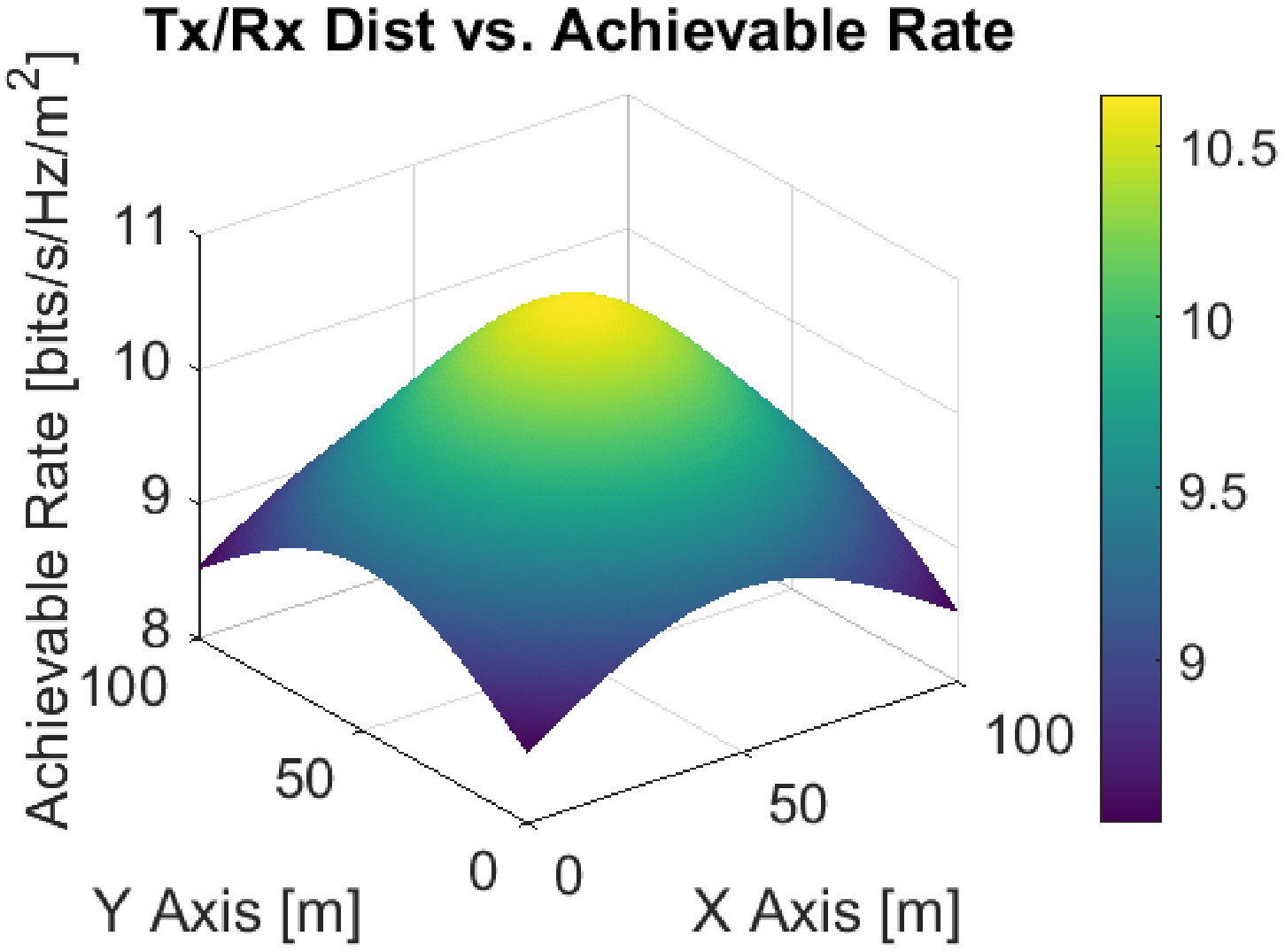}\par
(b)\par
\vspace{6pt}
Fig. 8. (a) Tx-Rx separation vs. achievable data rate (top view); (b) Tx-Rx separation vs. achievable data rate (3D)
\vspace{6pt}

\justifying{Fig. 9 illustrates the achievable rate in terms of the number of transmit-receive elements of IRS. $p_t$ = 2W (parameters are as same as Fig. 6).}
\vspace{6pt}

\centering
\includegraphics[height=8.0cm, width=10.0cm]{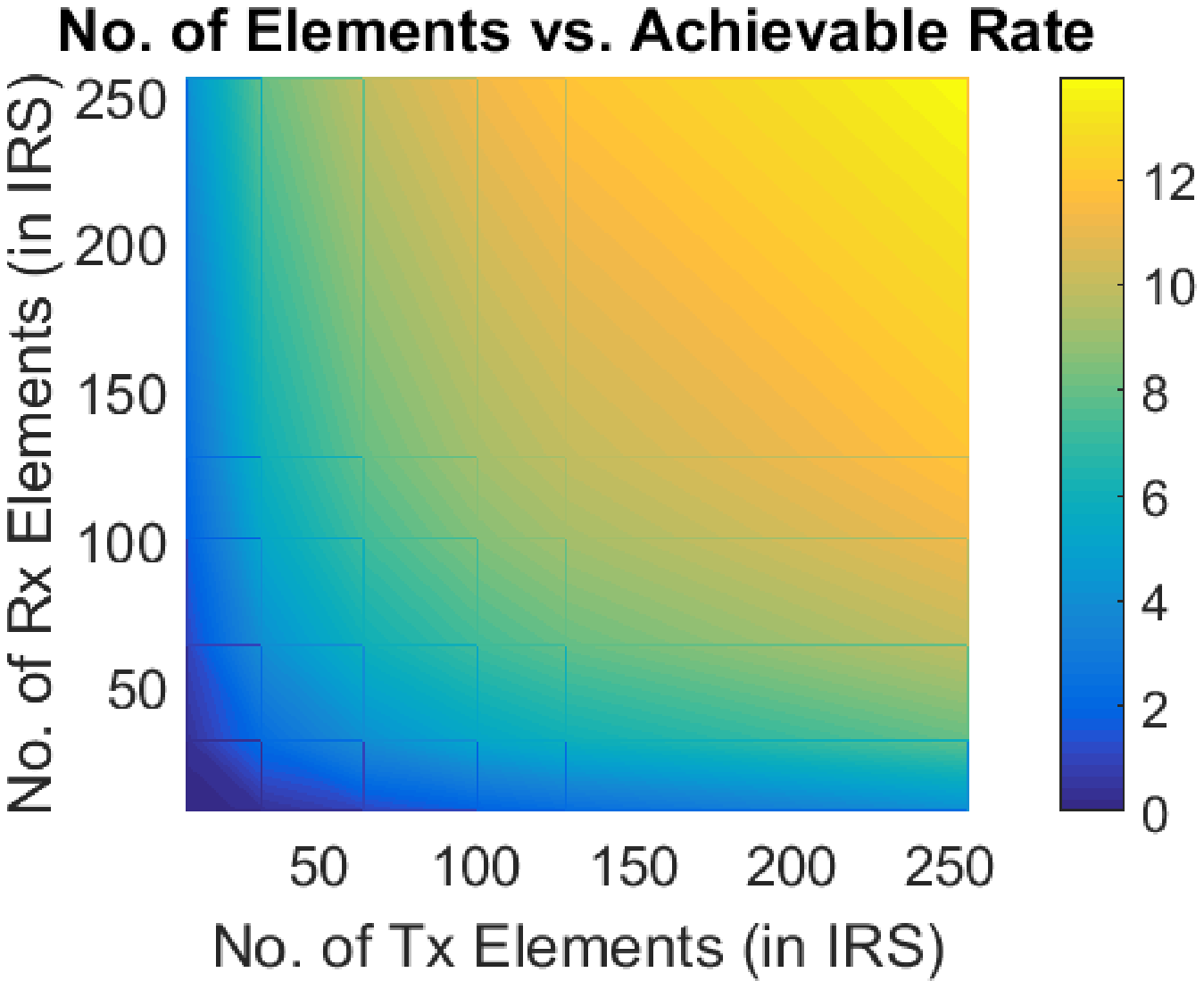}\par
\vspace{6pt}
Fig. 9. No. of Tx-Rx elements vs. achievable data rate
\vspace{6pt}

\justifying{Through the observation of Figs. 4 and 7, it is evident that $11\degree,$ $33\degree,$ $55\degree,$ $77\degree$  angles exhibits the highest path loss and lowest achievable data rate. $0\degree,$ $22\degree,$ $44\degree,$ $66\degree,$ $88\degree$  angles of transmissions and reception produce the lowest path loss and highest achievable data rate. But the path loss is quite lower and the achievable rate is much higher than the case of the conventional UAV communication model. According to the prior research $30\degree$ to $60\degree$ transmit-receive angle [51], [52] is prominent for IRS communication.\par
It can be interpreted by examining Figs. 5 and 8 that, with the increase of transmitter-receiver separation distance the path loss increases and the achievable data rate gradually decreases. Since the IRS enables almost perfect reflection of the transmitted signal, reduces the absorption, and establishes an LoS condition a greater number of users can obtain a significantly favorable signal (in the considered coverage region) compared to the conventional UAV communication model (comparing with Figs. 2 and 3).\par
According to the observation of Figs. 6 and 9 it is comprehensible that, with the increase of the number of transmit-receive elements in IRS the path loss decreases, and the achievable data rate increases gradually.\par
Observing and comparing the figures finally it is evident that, the deployment of IRS reduces the path loss by around 70 – 100 dB and increases the achievable data rate up to 8 – 13 times approximately (according to the chosen network parameter).\par
An intriguing observation of this research is that by utilizing a lower transmit power a significantly higher achievable data rate can be obtained by deploying IRS-empowered UAV communication networks (2W) compared to the conventional UAV communication model (6W).\par
The research considered low altitude UAVs since it utilized mmWave carrier to minimize the significant attenuation of the carrier. Moreover, the work considered a limited interference scenario.\par
Furthermore, the research considered IRS (UAV) 3D location in the reference formula [20] which makes the measurement more precise.}\par
\vspace{18pt}

\RaggedRight{\textbf{\Large 5.\hspace{10pt} Conclusion}}\\
\vspace{12pt}

\justifying{\noindent The research work targeted minimizing the path loss and maximizing the achievable data rate of conventional UAV-assisted communication by deploying IRS-empowered UAV-assisted wireless networks. The work, therefore, analyzed and compared the deployment of IRS-UAV communication and conventional UAV communication considering the aforementioned measurement model. Performing MATLAB-based measurements the research obtained that IRS-empowered UAV communication networks minimize the path loss and maximize the achievable data rate significantly. Moreover, the IRS-UAV communication model reduces the energy consumption of the entire network. Further research can be performed on the analyzed model considering higher mmWave and THz-band carriers, multiple UAVs, higher altitude, etc. The research will be assistive to the scholars, researchers, and enthusiasts who are engaged in the relative research.}\par
\vspace{18pt}

\RaggedRight{\textbf{\Large References}}\\
\vspace{12pt}
\begin{enumerate}
\justifying{
    \item Ahmed A. Ibrahim,  Jan Machac, and Raed M. Shubair. "Compact UWB MIMO antenna with pattern diversity and band rejection characteristics." Microwave and Optical Technology Letters 59, no. 6 (2017): 1460-1464.
    \item Raed M. Shubair and Hadeel Elayan. "In vivo wireless body communications: State-of-the-art and future directions." In 2015 Loughborough Antennas \& Propagation Conference (LAPC), pp. 1-5. IEEE, 2015.
    \item Hadeel Elayan, Raed M. Shubair, and Asimina Kiourti. "Wireless sensors for medical applications: Current status and future challenges." In 2017 11th European Conference on Antennas and Propagation (EUCAP), pp. 2478-2482. IEEE, 2017.
    \item Menna El Shorbagy, Raed M. Shubair, Mohamed I. AlHajri, and Nazih Khaddaj Mallat. "On the design of millimetre-wave antennas for 5G." In 2016 16th Mediterranean Microwave Symposium (MMS), pp. 1-4. IEEE, 2016.
    \item Hadeel Elayan, Raed M. Shubair, Josep Miquel Jornet, and Raj Mittra. "Multi-layer intrabody terahertz wave propagation model for nanobiosensing applications." Nano communication networks 14 (2017): 9-15.
    \item Hadeel Elayan, Pedram Johari, Raed M. Shubair, and Josep Miquel Jornet. "Photothermal modeling and analysis of intrabody terahertz nanoscale communication." IEEE transactions on nanobioscience 16, no. 8 (2017): 755-763.
    \item Rui Zhang, Ke Yang, Akram Alomainy, Qammer H. Abbasi, Khalid Qaraqe, and Raed M. Shubair. "Modelling of the terahertz communication channel for in-vivo nano-networks in the presence of noise." In 2016 16th Mediterranean Microwave Symposium (MMS), pp. 1-4. IEEE, 2016.
    \item M. I. AlHajri, A. Goian, M. Darweesh, R. AlMemari, R. M. Shubair, L. Weruaga, and A. R. Kulaib. "Hybrid RSS-DOA technique for enhanced WSN localization in a correlated environment." In 2015 International Conference on Information and Communication Technology Research (ICTRC), pp. 238-241. IEEE, 2015.
    \item Samar Elmeadawy and Raed M. Shubair. "6G wireless communications: Future technologies and research challenges." In 2019 international conference on electrical and computing technologies and applications (ICECTA), pp. 1-5. IEEE, 2019.
    \item Maryam AlNabooda, Raed M. Shubair, Nadeen R. Rishani, and Ghadah Aldabbagh. "Terahertz spectroscopy and imaging for the detection and identification of illicit drugs." 2017 Sensors networks smart and emerging technologies (SENSET) (2017): 1-4.
    \item Hadeel Elayan, Raed M. Shubair, and Asimina Kiourti. "On graphene-based THz plasmonic nano-antennas." In 2016 16th mediterranean microwave symposium (MMS), pp. 1-3. IEEE, 2016.
    \item M. Saeed Khan, A-D. Capobianco, Sajid M. Asif, Adnan Iftikhar, Benjamin D. Braaten, and Raed M. Shubair. "A pattern reconfigurable printed patch antenna." In 2016 IEEE International Symposium on Antennas and Propagation (APSURSI), pp. 2149-2150. IEEE, 2016.
    \item M. Saeeed Khan, Adnan Iftikhar, Sajid M. Asif, Antonio‐Daniele Capobianco, and Benjamin D. Braaten. "A compact four elements UWB MIMO antenna with on‐demand WLAN rejection." Microwave and Optical Technology Letters 58, no. 2 (2016): 270-276.
    \item M. A. Al-Nuaimi, R. M. Shubair, and K. O. Al-Midfa. "Direction of arrival estimation in wireless mobile communications using minimum variance distortionless response." In The Second International Conference on Innovations in Information Technology (IIT’05), pp. 1-5. 2005.
    \item Fahad Belhoul, Raed M. Shubair, and Mohammed E. Al-Mualla. "Modelling and performance analysis of DOA estimation in adaptive signal processing arrays." In ICECS, pp. 340-343. 2003.
    \item Muhammad Saeed Khan, Adnan Iftikhar, Antonio‐Daniele Capobianco, Raed M. Shubair, and Bilal Ijaz. "Pattern and frequency reconfiguration of patch antenna using PIN diodes." Microwave and Optical Technology Letters 59, no. 9 (2017): 2180-2185.
    \item Muhammad Saeed Khan, Adnan Iftikhar, Antonio‐Daniele Capobianco, Raed M. Shubair, and Bilal Ijaz. "Pattern and frequency reconfiguration of patch antenna using PIN diodes." Microwave and Optical Technology Letters 59, no. 9 (2017): 2180-2185.
    \item Ali Hakam, Raed M. Shubair, and Ehab Salahat. "Enhanced DOA estimation algorithms using MVDR and MUSIC." In 2013 International Conference on Current Trends in Information Technology (CTIT), pp. 172-176. IEEE, 2013.
    \item Hadeel Elayan, Cesare Stefanini, Raed M. Shubair, and Josep Miquel Jornet. "End-to-end noise model for intra-body terahertz nanoscale communication." IEEE transactions on nanobioscience 17, no. 4 (2018): 464-473.
    \item Ebrahim M. Al-Ardi, Raed M. Shubair, and Mohammed E. Al-Mualla. "Direction of arrival estimation in a multipath environment: An overview and a new contribution." Applied Computational Electromagnetics Society Journal 21, no. 3 (2006): 226.
    \item Mohamed I. AlHajri, Nazar T. Ali, and Raed M. Shubair. "Indoor localization for IoT using adaptive feature selection: A cascaded machine learning approach." IEEE Antennas and Wireless Propagation Letters 18, no. 11 (2019): 2306-2310.
    \item Hadeel Elayan, Raed M. Shubair, Akram Alomainy, and Ke Yang. "In-vivo terahertz em channel characterization for nano-communications in wbans." In 2016 IEEE International Symposium on Antennas and Propagation (APSURSI), pp. 979-980. IEEE, 2016.
    \item R. M. Shubair. "Robust adaptive beamforming using LMS algorithm with SMI initialization." In 2005 IEEE Antennas and Propagation Society International Symposium, vol. 4, pp. 2-5. IEEE, 2005.
    \item R. M. Shubair and A. Al-Merri. "Robust algorithms for direction finding and adaptive beamforming: performance and optimization." In The 2004 47th Midwest Symposium on Circuits and Systems, 2004. MWSCAS'04., vol. 2, pp. II-II. IEEE, 2004.
    \item Pradeep Kumar Singh, Bharat K. Bhargava, Marcin Paprzycki, Narottam Chand Kaushal, and Wei-Chiang Hong, eds. Handbook of wireless sensor networks: issues and challenges in current Scenario's. Vol. 1132. Berlin/Heidelberg, Germany: Springer, 2020.
    \item R. M. Shubair and Y. L. Chow. "A closed-form solution of vertical dipole antennas above a dielectric half-space." IEEE transactions on antennas and propagation 41, no. 12 (1993): 1737-1741.
    \item Ebrahim M. Al-Ardi, Raed M. Shubair, and Mohammed E. Al-Mualla. "Computationally efficient DOA estimation in a multipath environment using covariance differencing and iterative spatial smoothing." In 2005 IEEE International Symposium on Circuits and Systems, pp. 3805-3808. IEEE, 2005.
    \item R. M. Shubair and W. Jessmi. "Performance analysis of SMI adaptive beamforming arrays for smart antenna systems." In 2005 IEEE Antennas and Propagation Society International Symposium, vol. 1, pp. 311-314. IEEE, 2005.
    \item E. M. Al-Ardi, R. M. Shubair, and M. E. Al-Mualla. "Investigation of high-resolution DOA estimation algorithms for optimal performance of smart antenna systems." (2003): 460-464.
    \item E. M. Al-Ardi, Raed M. Shubair, and M. E. Al-Mualla. "Performance evaluation of direction finding algorithms for adapative antenna arrays." In 10th IEEE International Conference on Electronics, Circuits and Systems, 2003. ICECS 2003. Proceedings of the 2003, vol. 2, pp. 735-738. IEEE, 2003.
    \item S. Gong et al., "Toward Smart Wireless Communications via Intelligent Reflecting Surfaces: A Contemporary Survey," in IEEE Communications Surveys \& Tutorials, vol. 22, no. 4, pp. 2283-2314, Fourthquarter 2020.
    \item M. Di Renzo et al., "Smart Radio Environments Empowered by Reconfigurable Intelligent Surfaces: How It Works, State of Research, and The Road Ahead," in IEEE Journal on Selected Areas in Communications, vol. 38, no. 11, pp. 2450-2525, Nov. 2020.
    \item H. Elayan, O. Amin, B. Shihada, R. M. Shubair and M. -S. Alouini, "Terahertz Band: The Last Piece of RF Spectrum Puzzle for Communication Systems," in IEEE Open Journal of the Communications Society, vol. 1, pp. 1-32, 2020.
    \item A. Masaracchia et al., "UAV-Enabled Ultra-Reliable Low-Latency Communications for 6G: A Comprehensive Survey," in IEEE Access, vol. 9, pp. 137338-137352, 2021.
    \item C. Pan et al., "Reconfigurable Intelligent Surfaces for 6G Systems: Principles, Applications, and Research Directions," in IEEE Communications Magazine, vol. 59, no. 6, pp. 14-20, June 2021.
    \item Q. Wu, S. Zhang, B. Zheng, C. You and R. Zhang, "Intelligent Reflecting Surface-Aided Wireless Communications: A Tutorial," in IEEE Transactions on Communications, vol. 69, no. 5, pp. 3313-3351, May 2021.
    \item S. Zhang et al., "Intelligent Omni-Surfaces: Ubiquitous Wireless Transmission by Reflective-Refractive Metasurfaces," in IEEE Transactions on Wireless Communications, vol. 21, no. 1, pp. 219-233, Jan. 2022.
    \item A. A. Khuwaja, Y. Chen, N. Zhao, M. -S. Alouini and P. Dobbins, "A Survey of Channel Modeling for UAV Communications," in IEEE Communications Surveys \& Tutorials, vol. 20, no. 4, pp. 2804-2821, Fourthquarter 2018, doi: 10.1109/COMST.2018.2856587.
    \item X. Pei et al., "RIS-Aided Wireless Communications: Prototyping, Adaptive Beamforming, and Indoor/Outdoor Field Trials," in IEEE Transactions on Communications, vol. 69, no. 12, pp. 8627-8640, Dec. 2021.
    \item X. Pang, M. Sheng, N. Zhao, J. Tang, D. Niyato and K. -K. Wong, "When UAV Meets IRS: Expanding Air-Ground Networks via Passive Reflection," in IEEE Wireless Communications, vol. 28, no. 5, pp. 164-170, October 2021.
    \item D. Ma, M. Ding and M. Hassan, "Enhancing Cellular Communications for UAVs via Intelligent Reflective Surface," 2020 IEEE Wireless Communications and Networking Conference (WCNC), 2020, pp. 1-6.
    \item M. Al-Jarrah, A. Al-Dweik, E. Alsusa, Y. Iraqi and M. S. Alouini, "On the Performance of IRS-Assisted Multi-Layer UAV Communications With Imperfect Phase Compensation," in IEEE Transactions on Communications, vol. 69, no. 12, pp. 8551-8568, Dec. 2021.
    \item X. Pang, M. Sheng, N. Zhao, J. Tang, D. Niyato and K. -K. Wong, "When UAV Meets IRS: Expanding Air-Ground Networks via Passive Reflection," in IEEE Wireless Communications, vol. 28, no. 5, pp. 164-170, October 2021.
    \item H. Jiang, R. He, C. Ruan, J. Zhou and D. Chang, "Three-Dimensional Geometry-Based Stochastic Channel Modeling for Intelligent Reflecting Surface-Assisted UAV MIMO Communications," in IEEE Wireless Communications Letters, vol. 10, no. 12, pp. 2727-2731, Dec. 2021.
    \item Z. Wei et al., "Sum-Rate Maximization for IRS-Assisted UAV OFDMA Communication Systems," in IEEE Transactions on Wireless Communications, vol. 20, no. 4, pp. 2530-2550, April 2021.
    \item A. Mahmoud, S. Muhaidat, P. C. Sofotasios, I. Abualhaol, O. A. Dobre and H. Yanikomeroglu, "Intelligent Reflecting Surfaces Assisted UAV Communications for IoT Networks: Performance Analysis," in IEEE Transactions on Green Communications and Networking, vol. 5, no. 3, pp. 1029-1040, Sept. 2021.
    \item C. Cao, Z. Lian, Y. Wang, Y. Su and B. Jin, "A Non-Stationary Geometry-Based Channel Model for IRS-Assisted UAV-MIMO Channels," in IEEE Communications Letters, vol. 25, no. 12, pp. 3760-3764, Dec. 2021.
    \item T. Shafique, H. Tabassum and E. Hossain, "Optimization of Wireless Relaying With Flexible UAV-Borne Reflecting Surfaces," in IEEE Transactions on Communications, vol. 69, no. 1, pp. 309-325, Jan. 2021.
    \item M. Mozaffari, W. Saad, M. Bennis and M. Debbah, "Optimal Transport Theory for Cell Association in UAV-Enabled Cellular Networks," in IEEE Communications Letters, vol. 21, no. 9, pp. 2053-2056, Sept. 2017.
    \item W. Tang et al., “Path Loss Modeling and Measurements for Reconfigurable Intelligent Surfaces in the Millimeter-Wave Frequency Band,” 2021, arXiv: 2101.08607v2. [Online]. Available: https://arxiv.org/abs/2101.08607
    \item W. Tang et al., "Wireless Communications With Reconfigurable Intelligent Surface: Path Loss Modeling and Experimental Measurement," in IEEE Transactions on Wireless Communications, vol. 20, no. 1, pp. 421-439, Jan. 2021.
    \item Ö. Özdogan, E. Björnson and E. G. Larsson, "Intelligent Reflecting Surfaces: Physics, Propagation, and Pathloss Modeling," in IEEE Wireless Communications Letters, vol. 9, no. 5, pp. 581-585, May 2020.
    }
\end{enumerate}

\end{document}